\documentclass[12pt]{article}

\usepackage{bbm}
\usepackage{t1enc}
\usepackage[latin2]{inputenc}
\usepackage{amsmath,amssymb,amsthm}
\usepackage{txfonts}
\usepackage{sectsty}
\usepackage{overcite}

\def\Tr{\text{Tr }}

\newcommand{\eq}[1]{\begin{equation}#1\end{equation}}
\newcommand{\define}{\coloneqq}
\newcommand{\affiliation}[1]{
              \noindent\hspace*{2.5em}\parbox[t]{10cm}{\small\it#1}
              \vspace*{2ex}\\}

\theoremstyle{plain}\newtheorem*{theor}{Theorem}
\theoremstyle{plain}\newtheorem*{prop}{Proposition}

\renewcommand{\leq}{\leqslant}
\renewcommand{\le}{\leqslant}
\renewcommand{\ge}{\geqslant}
\renewcommand{\geq}{\geqslant}

\renewcommand{\title}[1]{{\Large\bf\flushleft{#1}}\vspace*{3ex}\\}
\renewcommand{\author}[2]{{\noindent\hspace*{2.5em}\large#1}
                     \footnote{Electronic mail: $\mathtt{#2}$}\\}

\sectionfont{\normalsize\bf}
\subsectionfont{\normalsize\bf}
\makeatletter\renewcommand\@biblabel[1]{$^{#1}$}\makeatother

\begin{document}

 \title{On the sharpness of the zero-entropy-density conjecture}

\author{S.~Farkas}{farkas@uchicago.edu}
\affiliation{Department of Physics, University of Chicago, Chicago,\\
             Illinois 60637}

\author{Z.~Zimbor\'as}{cimbi@rmki.kfki.hu} 
\affiliation{Research Institute for Particle and Nuclear Physics, Budapest,\\
             H-1525 Budapest, P.O. Box 49, Hungary}

\setcounter{footnote}{0}
\vspace*{-3ex}
\begin{abstract}
The zero-entropy-density conjecture states that the entropy density defined as 
\mbox{$s\define\lim_{N\to\infty}S_N/N$} vanishes for all translation-invariant 
pure states on the spin chain. Or equivalently, $S_N$, the von Neumann entropy
of such a state restricted to $N$ consecutive spins, is sublinear. 
In this paper it is proved that this conjecture cannot be sharpened, 
i.e., translation-invariant states give rise to arbitrary fast sublinear
entropy growth. The proof is constructive, and is based on a 
class of states derived from quasifree states on a CAR algebra. 
The question whether the entropy growth of pure quasifree states can be 
arbitrary fast sublinear was first raised by  Fannes {\it et al.}
[J. Math Phys. {\bf 44}, 6005 (2003)]. In addition to the main theorem 
it is also shown that the entropy asymptotics of all pure
shift-invariant nontrivial quasifree states is at least logarithmic.
\end{abstract}

\section{INTRODUCTION}

\hspace{\parindent}Quantum spin chains belong to the most studied models
of quantum statistical mechanics.\cite{Bratteli} 
Still, only for a few types of models have the thermal 
and ground state structures been determined.
This is mainly the consequence of the complicated correlations
that can appear in quantum states.
These strong correlations can even be
present in pure states,
while classical pure states 
can only 
have a
trivial product state structure.  
Unlike the classical case, the restrictions
of pure states on the quantum spin chain to 
local subsystem are typically mixed states. 
  This type of 
correlation between  
subsystems
is commonly referred to as entanglement.
The von Neumann entropy, defined as 
\mbox{$S\!\coloneqq\!-\Tr \rho \log \rho$}, is a 
natural measure of the nonpurity of the restricted density matrix $\rho$, 
thus it is a very useful quantity in the description of entanglement. 

The entropy of a restricted density matrix
is also a basic measure when
mixed states are treated, however, in this case it
cannot be interpreted as a measure of entanglement.
Let $S_{N}$ denote the von Neumann entropy 
of a translation-invariant state restricted to $N$ consecutive spins. 
The entropy density $s\!\coloneqq\!\lim_{N \to\infty} S_N/N$ is 
considered to quantify the "strong nonpurity" of the
entire mixed state, and it plays a central role in the characterization of
Gibbs states. \cite{Bratteli}   
A natural and long-standing conjecture 
is that the entropy density vanishes for all
translation-invariant pure states on a quantum spin chain,
i.e., for such states, $S_{N}$ is sublinear.
In the present paper we will prove that if this 
zero-entropy-density conjecture is true, 
then it is sharp in the sense that for any sublinear function $f_{N}$
$(\lim_{N \to\infty} f_N/N=0)$,
there exists a translation-invariant state so that 
$S_{N}\!\ge\!f_{N}$, for every sufficiently large $N$. This has already
been conjectured 
by Fannes, Haegeman, and Mosonyi in \mbox{Ref. \citeonline{mosonyi}}.
Moreover, they proved that any sublinear power function can be exceeded
by the entropy growth of an appropriate pure translation-invariant state. 
 
It also should be mentioned that
there is a revived interest in studying 
entropy asymptotics for two other reasons.
First, $S_{N}$ seems to be a good indicator of
quantum criticality. Several ground states
of Hamiltonians with local interactions were studied,
and in these models,
$S_{N}$ was found to be bounded for
noncritical systems, while for critical systems,
it turned out to diverge logarithmically.\cite{Vidal,Korepin,Keating,Viktor}
The prefactor of this logarithmic growth was argued to be
one-third of the central charge.\cite{Cardy}
Also higher dimensional lattice models have been investigated in this
respect.\cite{Eisert1,Eisert2,Wolf,Gioev} Second, entropy  is supposed to
play an important role in
the quantification of the "essential subspace"
of a restricted density matrix.
The possibility of compressing the restricted
density matrix from its full dimension to a
much smaller subspace without loss
of much information is the starting point of
the DMRG
calculations.\cite{White} For ergodic translation-invariant states,
with non-vanishing entropy density $s$,
the density matrix pertaining to $N$ consecutive spins, $\rho_{N}$, 
is essentially concentrated on a subspace of
dimension proportional to $\exp(Ns)$.\cite{Petz}
In more general situations, numerical calculations suggest 
that the dimension of the "essential subspace" of $\rho_{N}$ is
proportional to $\exp(S_{N})$.\cite{Legeza} This could lead to a very
efficient compression of states with
bounded or slowly diverging entropy asymptotics. 

In the present paper we give a constructive proof 
of the sharpness of the zero-entropy-density conjecture.
The states that are studied are 
translation-invariant pure states on the 
spin chain derived from 
quasifree states on the CAR algebra. In Sec. II
we recapitulate the construction of such states in 
order to be self-contained. In Sec. III
we prove our main theorem.
The argument is based on the approach to quasifree states
developed in \mbox{Ref. \citeonline{mosonyi}}.
Finally, we include a proof
of the statement that
the entropy growth for
all nontrivial gauge-invariant quasifree states are bounded from below by a 
logarithmic growth.

\section{QUASIFREE STATES ON THE SPIN CHAIN} \label{qf}

\subsection{The Araki-Jordan-Wigner construction}

\hspace{\parindent}The algebra of observables of a 
quantum spin chain\footnote{More precisely,
spin $\frac{1}{2}$ chain.}
is the UHF algebra
\begin{equation*}
\mathcal{M}:= \bigotimes^{+\infty}_{k=-\infty}M_{2},
\end{equation*}
where $M_{2}$ denotes the algebra of $2 \times 2$ matrices.
Let $\sigma_a^k$ $(a=1,2,3;\;k\in\mathbb{Z})$ denote the Pauli matrices 
embedded into the $k$th factor of $\mathcal{M}$. They satisfy the 
well-known relations:
\begin{eqnarray*}
\sigma_a^k\sigma_b^l &=&\sigma_b^l\sigma_a^k,\quad \text{when} \quad k \ne l,\\
\sigma_a^l\sigma_b^l &=& \mathrm{i}\varepsilon_{abc}\sigma_c^l+\delta_{ab}
\mathbbm{1}.
\end{eqnarray*}
The Pauli matrices and $\mathbbm{1}$ generate $\mathcal{M}$. 
The translation automorphism $\tau$
on $\mathcal{M}$ is defined by 
$\tau(\sigma^k_a)=\sigma^{k+1}_a$. 

The states we are investigating in this paper are translation-invariant
pure states on $\mathcal{M}$ derived from quasifree states of
a fermion chain.
The $C^{*}$ algebra 
describing a fermion chain
is the CAR algebra corresponding to the one-particle Hilbert space 
$\ell^{2}(\mathbb{Z})$, 
i.e. it is the $C^{*}$ algebra generated by
$\mathbbm{1}$ and the operators $\{ c_{k} \bigm| k \in \mathbb{Z} \}$
satisfying the canonical anticommutation relations:
\begin{eqnarray*}
      c_{k}c^*_{l}+c^*_{l}c_{k}=\delta_{k,l} \mathbbm{1}, \quad \quad
      c_{k}c_{l}+c_{l}c_{k}=0.
\end{eqnarray*}
Denote this $C^{*}$ algebra by $\mathcal{A}$.
The translation automorphism $\gamma$ 
is defined by $\gamma(c_{k})= c_{k+1}$. 

The $C^{*}$ algebras $\mathcal{M}$ and $\mathcal{A}$
are isomorphic.
However, there
exists no isomorphism
$\iota \colon \mathcal{M} \to \mathcal{A}$ that satisfies 
the property $\iota \circ \tau = \gamma \circ \iota $.\footnote{This is 
clear if we note that $(\mathcal{M},\tau)$ is asymptotically
 Abelian, while $(\mathcal{A},\gamma)$ is not.} 
This intertwining property is needed 
to derive the translation invariance of a state $\omega\circ\iota$ on 
$\mathcal{M}$ from that of $\omega$ on $\mathcal{A}$. 
This problem can be circumvented by Araki's construction. 
\cite{Araki1} 
In this section we will present a modified but equivalent
formulation of this construction.

First, let us introduce the parity automorphism $\pi$ on $\mathcal{A}$. 
It is defined by {$\pi(c_{k})\!=\!-c_{k}$.} 
The elements of \mbox{$\mathcal{A}_{+}:=\{a\in\mathcal{A}\bigm|\pi(a)=a\}$} 
are called even,
while those of \mbox{$\mathcal{A}_{-}:=\{a\in\mathcal{A}\bigm|\pi(a)=-a\}$} 
are called odd. Any element $a \in \mathcal{A}$
can uniquely be written in the form 
$a\!=\!a_+ + a_-$, where $a_+\!\in\!\mathcal{A}_+$, and $a_-\!\in\!\mathcal{A}_-$. 
Thus, $\mathcal{A}=\mathcal{A}_{+}+\mathcal{A}_{-}$.

Second, let $\mathcal{M}_+$ be the $C^*$ subalgebra of $\mathcal{M}$ 
generated by $\mathbbm{1}$, $\sigma_3^k$, and $\sigma_1^k\sigma_1^l$ 
\mbox{$(k,l\in\mathbb{Z})$}. $\mathcal{M}_+$ is isomorphic to $\mathcal{A}_+$,
an explicit isomorphism $\alpha$ is given by the restricted
Jordan-Wigner transformation:
\begin{eqnarray*}
\alpha(\sigma_3^k)&:=&2c_k^* c_k - \mathbbm{1}, \\
\alpha(\sigma_1^k \sigma_1^l)&:=&-\prod_{m=k}^{l-1} (2c_m^* c_m -
\mathbbm{1})
(c_k^*+c_k)(c_l^*+c_l) \quad \text{when} \quad k < l.  
\end{eqnarray*} 

Since $\mathcal{M}_+$ and $\mathcal{A}_+$ are invariant under the
translations, $\tau$ and $\gamma$ can be restricted to
$\mathcal{M}_+$ and $\mathcal{A}_+$, respectively. 
Let us denote these 
restrictions by $\tau_+$ and $\gamma_+$. 
Although there is no isomorphism that intertwines the translations on 
$\mathcal{M}$ and $\mathcal{A}$, $\alpha$ is an isomorphism that 
intertwines the translations on  
the subalgebras $\mathcal{M}_+$ and 
$\mathcal{A}_+:\alpha\circ\tau_+=\gamma_+\circ\alpha$.

Now, let $\omega_+$ be the restriction of a state
 $\omega$ on $\mathcal{A}$ to $\mathcal{A}_+$. 
If $\omega$ is a translation-invariant 
state, i.e., $\omega\circ\gamma=\omega$, then $\omega_+\circ\alpha$, which is 
a state on $\mathcal{M}_+$, 
is invariant under $\tau_+$. The state $\omega_+\circ\alpha$ 
can be extended to a state
 $\widetilde{\omega}$ on $\mathcal{M}$ by 
$\widetilde{\omega}(a)=\widetilde{\omega}(a_++a_-):=\omega_+(\alpha(a_+))$,
 where $a_+\in\mathcal{A}_+$, and $a_-\in\mathcal{A}_-$.
This way a translation-invariant state $\widetilde{\omega}$ on 
$\mathcal{M}$ is obtained. Moreover, if $\omega$ is an 
even state, i.e., $\omega\circ\pi=\omega$, then $\widetilde{\omega}$ is pure 
if and only if $\omega$ is pure. \cite{mosonyi}

To summarize, a translation intertwining automorphism $\alpha$ has been given
not between the algebras $\mathcal{M}$ and $\mathcal{A}$ but between their 
appropriate subalgebras $\mathcal{M}_+$ and $\mathcal{A}_+$. Any 
translation-invariant state on $\mathcal{M}_+$ can be straightforwardly 
extended to a translation-invariant state on $\mathcal{M}$. Thus the 
isomorphism $\alpha$ makes it possible to transport the translation-invariant
states from $\mathcal{A}$ to $\mathcal{M}$.

\subsection{Quasifree states on CAR algebras}

\hspace{\parindent}Following Araki's construction 
presented in the previous section,
a class of states will be derived from quasifree states on the 
CAR algebra $\mathcal{A}$. In this section we will shortly recapitulate
the most important definitions and facts concerning these states, more details
and the proofs of the statements can be found in Ref. \citeonline{Bratteli} and
Ref. \citeonline{Alicki}.

Let $Q$ be an operator on the Hilbert space $\ell^{2}(\mathbb{Z})$,
$0\!\le\!Q\!\le\!\mathbbm{1}$. 
Let $Q_{ij}:=\langle\delta_{i}, Q \delta_{j}\rangle$ be the matrix elements of 
$Q$ in the standard basis 
$\{ \delta_{k} \bigm| k \in \mathbb{Z} \}$ of $\ell^{2}(\mathbb{Z})$,
where $\delta_{k}$ is the characteristic function of the number $k$.
The (gauge-invariant)
quasifree state $\omega_{Q}$ on $\mathcal{A}$ is
defined through the following formula:
\begin{equation*}
\omega_{Q}(c_{i_{1}}^{*} \dots c_{i_{m}}^{*}c_{j_{n}} \dots c_{j_{1}})
=\delta_{m,n} 
\text{det} \left( [Q_{i_{k}j_{l}}]_{k,l=1}^{n} \right),
\end{equation*} 
The operator $Q$ is called the symbol of the state. Quasifree states
are by definition even states.

A quasifree state $\omega_{Q}$ is translation-invariant if and only if its
symbol $Q$ is a Toeplitz operator in the basis $\delta_{k}$,
i.e., there exists a sequence $(q_k)_{k\in\mathbb{Z}}$
such that $Q_{kl}=\langle\delta_{k}, Q \delta_{l}\rangle=q_{k-l}$.
Let us introduce the Fourier transform:
\[
\tilde{q}(\theta):=\sum_{k \in \mathbb{Z}}q_{k}e^{i 2 \pi k \theta},
\quad\quad \text{where } \theta\in[0,1).
\] 
The function $\tilde{q}$ satisfies $0 \le \tilde{q}(\theta) \le 1$ 
almost everywhere. A translation-invariant 
quasifree state $\omega_{Q}$ 
is pure if and only if the Fourier transform $\tilde{q}$ is a 
characteristic function, 
i.e., there exists a measurable set $K \subset[0,1)$ such that 
$\tilde{q}(\theta)=\chi_{K}(\theta)$.

Now, applying the Araki-Jordan-Wigner construction to a
translation-invariant quasifree state $\omega_{Q}$, 
one obtains a translation-invariant state $\widetilde{\omega}_{Q}$ on
the spin chain algebra $\mathcal{M}$. Since quasifree states are even,
the state $\widetilde{\omega}_{Q}$ is pure if and only if the corresponding 
quasifree state $\omega_{Q}$ is also pure. 

Let $\rho_{N}$ denote the reduced density matrix obtained
by restricting the state $\widetilde{\omega}_{Q}$ to an
interval of $N$ spins. The von Neumann entropy of the restricted state
is defined as \mbox{$S_{N}:=- \Tr \rho_{N} \log \rho_{N}$.}
An explicit formula of this entropy is known for quasifree states: 
\begin{equation}
S_{N}=-  \Tr \left( Q_{N}\log Q_{N} +(\mathbbm{1}-Q_{N}) 
\log (\mathbbm{1}-Q_{N})\right), 
\label{entropy}
\end{equation}
where $Q_{N}$ is the restriction of $Q$ to the $N$-dimensional
space spanned by the set $\{\delta_{k}\bigm| 0\le k\le N-1 \}$. 

On the basis of the Szeg\H{o} theorem one can prove that
the entropy density $s:= \lim_{N\to\infty}S_N/N$
of pure translation-invariant quasifree states vanishes.\cite{mosonyi}
In the next section we will prove that this
statement is sharp in the sense that for any $f_N$ sublinear
function there is a quasifree state for which
$S_{N} \ge f_{N}$ for sufficiently large $N$.

\section{QUASIFREE STATES GIVE RISE TO ARBITRARY FAST SUBLINEAR ENTROPY 
         GROWTH}
\label{th}

\hspace{\parindent}An explicit formula of the entropy 
function $S_N$ for quasifree states was 
given in the previous section by equation \eqref{entropy}. 
In order to simplify further computations, we work with a quadratic 
lower bound of $S_N$ introduced in Ref. \citeonline{mosonyi}:
\[q_N:=\mathop{\mathrm{Tr}}Q_N\left(\mathbbm{1}-Q_N\right)\label{qN_def}\]
That $q_N$ is a lower bound of 
$S_N=-\mathop{\mathrm{Tr}}(Q_N\log Q_N+
(\mathbbm{1}-Q_N)\log(\mathbbm{1}-Q_N))$ 
can be proved by the aid of the inequality
 $x(1-x)\leq-x\ln x-(1-x)\ln(1-x)$, which holds for $0\leq x\leq 1$.

As derived in Ref. \citeonline{mosonyi}, $q_N$ can be rewritten in the form:
\eq{q_N=\int\limits_0^1d\phi\frac{\sin^2N\pi\phi}{\sin^2\pi\phi}
\Lambda_K(\phi),\label{qN}}
where $K$ denotes the measurable set $K$ that characterizes the symbol
 $Q$, and $\Lambda_K$ is the function:
\[\Lambda_K(\phi)=|(K+\phi)\setminus K|.\label{LK_def}\]
$|\cdot|$ denotes the Lebesgue measure. By reducing the region 
of the integration in  \eqref{qN} to $[0,1/(2N)]$, and substituting 
the trigonometric factor with its lower bound on this restricted region, 
we obtain a lower estimate for $q_N$:
\eq{q_N\geq\frac{4N^2}{\pi^2}\int\limits_0^\frac{1}{2N}\Lambda_K(\phi)d\phi.
\label{qN_start}}
This is the starting point in the proof of the following proposition.
\begin{theor} For any sublinear function 
$f\colon\mathbb{N}\to\mathbb{R}^+$, there exists a 
pure translation-invariant quasifree state for which 
$S_N$ is bounded from below by $f_N$, that is \mbox{$S_N\geq f_N$} 
for every sufficiently large  $N$.
\end{theor}
\begin{proof}
By (\ref{qN_start}), the problem has been reduced to showing the 
existence of a set $K\!\subset\![0,1)$ for which the right hand side of 
(\ref{qN_start}) grows not slower than the given $f_N$ as $N$ goes to infinity.

The construction of $K$ is based on two non-negative sequences: a sequence of 
integers
$(n_i)_{i\in\mathbb{N}}$ and another one of real numbers $
(\ell_i)_{i\in\mathbb{N}}$, 
where $\ell_i\geq 2\ell_{i+1}$. 
Let $K$ be the union of infinitely many disjoint intervals, 
the end points of which are determined by these two sequences as follows:
\eq{
\begin{split}
K&=\bigcup_{i\in\mathbb{N}}\;
\bigcup_{k=1}^{n_i}I_i^k,\;\;\;I_i^k=[a_i^k,b_i^k],\;\;\;b_i^k-a_i^k=\ell_i;\\
a_0^1=0;\;\;\;\;\;&a_i^1=b_{i-1}^{n_{i-1}}+\ell_{i-1},
\;\;\;\mbox{if}\;i>0;\label{construction}\\
&a^k_i=b^{k-1}_i+\ell_i,\;\;\;\;\;\;\mbox{if}\;k>1.
\end{split}
}
The $(\ell_i)_{i\in\mathbb{N}}$ and $(n_i)_{i\in\mathbb{N}}$ are chosen 
so that the set $K$ above-constructed is bounded, and for convenience, 
we suppose additionally that $n_i\ell_i$ is monotonically decreasing, and:
\[\sum_{i=0}^\infty n_i\ell_i<\frac{1}{4}.\label{Kbounded}\]
Thus $K\subset[0,1/2]$. 
With construction \eqref{construction}, $\Lambda_K$ takes the form
\[
\Lambda_K(\phi)=\sum_{i=0}^\infty\sum_{k=1}^{n_i}
\left|(I_i^k+\phi)\setminus K\right|\geq\sum_{i=i_\phi}^\infty
\sum_{k=1}^{n_i}\left|(I_i^k+\phi)\setminus K\right|,
\]
where $i_{\phi}$ is the smallest index for which $2n_i\ell_i<\phi$ 
for all $i\geq i_\phi$. Each translated interval $(I_i^k+\phi)$ 
with $i\geq i_\phi$ is situated in a region where the original intervals  
in the construction of $K$ and the gaps between them 
are not longer than $\ell_i/2$ 
(or where $K$ has no point at all).    
For this reason $|(I_i^k+\phi)\setminus K|\geq \ell_i/3$ for every term in the 
last summation. Therefore we obtain
\[\Lambda_K(\phi)\geq\frac{1}{3}\sum_{i=i_\phi}^\infty n_i\ell_i.\]
Now, let $f_N$ be an arbitrary sublinear function, i.e., 
$\lim_{N\to\infty} f_N/N=0$. Obviously, there exists a monotonically 
increasing 
continuously differentiable function 
$g\colon [0,1/2]\to \mathbb{R}^+$ with the properties:  
\[
g(0)=0,\;\;\;\;\;\frac{\pi^2}{2}g\left(\frac{1}{2N}\right)\geq\frac{f_N}{N}.
\]
Let us define function 
$h$ as $h(x)=\frac{d}{dx}(xg(x))$. $h$ is continuous,
and $h(0)=0$. We suppose that $h$ is strictly monotonically increasing in the 
neighborhood of zero. If not, we choose a continuous, strictly monotonically 
increasing $\hat{h}$ 
such that $\hat{h}\ge h$, and $\hat{h}(0)=0$.\footnote{A possible choice is
$\hat{h}(x)\coloneqq\max\{h(y)\mid y\in[0,x]\}+x$.} 
This $\hat{h}$ can be derived 
from a $\hat{g}$ for which $\hat{g}\ge g$, and then the argument can be 
continued with $\hat{h}$ instead of $h$.  

The next step is to specify 
$(n_i)_{i\in\mathbb{N}}$ and $(\ell_i)_{i\in\mathbb{N}}$ so that   
\eq{
\label{l>h}
\Lambda_K(\phi)\geq\frac{1}{3}\sum_{i=i_\phi}^\infty n_i \ell_i
\geq h(\phi)
}
should hold for sufficiently small $\phi$. 

Let $s_i$ be the solution of the following recursive equation,
starting from a given $s_0$ (\mbox{$0<s_0<1/2$}):
\eq{
\label{seq}
h\left(6(s_i-s_{i+1})\right)=s_{i+1}.
}
It is clear from the required properties of $h$ that there is a solution that 
satisfies the equalities $0\leq s_{i+1}\leq s_i$ for every $i$. 
Since $(s_i)_{i\in\mathbb{N}}$ is bounded from below and monotonically 
decreasing, it has a limit at infinity. 
Suppose that this limit differs from zero, 
say it is $s_\infty>0$. Taking an arbitrary small $\epsilon>0$, 
there is an $i$ for which $\epsilon>6(s_i-s_{i+1})$, 
and we find that $h(\epsilon)\geq h(6(s_i-s_{i+1}))=s_{i+1}\!\geq\! s_\infty$ 
for any $\epsilon$, so $h(0)\geq s_\infty$ in contradiction with $h(0)=0$. 
Thus $\lim\nolimits_{i\to\infty}s_i=0$.

Now we are ready to specify the values of $\ell_i$ and $n_i$ by
\eq{
\label{nlfroms}
s_i=\frac{1}{3}\sum_{j=i}^\infty n_j\ell_j
}
Considering that $(s_i)_{i\in\mathbb{N}}$ is a monotonically 
decreasing sequence
tending to zero, these equalities can be satisfied by some series
$(n_i)_{i\in\mathbb{N}}$ and $(\ell_i)_{i\in\mathbb{N}}$. 
Starting with a particular $\ell_i$, 
we can always determine the next term by choosing some $\ell_{i+1}\le\ell_i/2$. 
The only restriction on the choice of $\ell_i$ is that $s_i-s_{i+1}$ should 
be an integral multiple of $\ell_i$. 
This requirement can undoubtedly be met, and then $s_i-s_{i+1}=\frac{1}{3}
n_i\ell_i$ yields the value of $n_i$. The inclusion $K\subset[0,1/2]$ 
can be assured by choosing sufficiently small $s_0$.         

Recall that $(n_i\ell_i)_{i\in\mathbb{N}}$ has been required to be monotonic. 
We can easily convince ourselves that $(n_i\ell_i)_{i\in\mathbb{N}}$ 
constructed from $(s_i)_{i\in\mathbb{N}}$ has this property. 
Indeed, it follows immediately from the strict monotonicity of $h$: 
$h(2n_i\ell_i)=h(6(s_i-s_{i+1}))=s_{i+1}\le s_i=h(6(s_{i-1}-s_i))
=h(2n_{i-1}\ell_{i-1})$.

Monotonicity of $(s_i)_{i\in\mathbb{N}}$ and its behavior 
at infinity entail that for any 
$\phi$ below a certain bound, there is an index $i$ for which 
$6(s_i-s_{i+1})\leq\phi\leq 6(s_{i-1}-s_i)$. 
Notice that this index is nothing but 
$i_\phi$. 
Thus putting together \eqref{seq}, and \eqref{nlfroms}, 
we arrive at the desired estimate \eqref{l>h}. Consequently, for sufficiently
large $N$, in the region of the integration in \eqref{qN_start}, 
$\Lambda_K(\phi)\geq h(\phi)$ holds.  
Performing the integration in \eqref{qN_start} completes the proof:
\begin{eqnarray*}
\lefteqn{S_N\ge q_N\geq\frac{4N^2}{\pi^2}\int\limits_0^\frac{1}{2N}
\Lambda_K(\phi)d\phi\geq\frac{4N^2}{\pi^2}\int\limits_0^\frac{1}{2N}
h(\phi)d\phi=}\\&&\frac{4N^2}{\pi^2}\int\limits_0^\frac{1}{2N}
\frac{d}{d\phi}\big(\phi g(\phi)\big)d\phi=\frac{2N}{\pi^2}g
\left(\frac{1}{2N}\right)\geq f_N.
\end{eqnarray*}         
\end{proof}  

We have just shown that pure translation-invariant quasifree states 
give rise to 
arbitrary fast sublinear entropy growth. In the trivial cases, $|K|=0$ 
or $|K|=1$, the entropy is identically zero. The question naturally arises 
whether it is possible to achieve arbitrary 
slow nonbounded entropy growth by such states. 
\begin{prop}
Apart from the trivial cases, pure (gauge- and) translation-invariant 
quasifree states give at least logarithmic entropy growth.
\end{prop}
\begin{proof}
It has been shown in Ref. \citeonline{mosonyi} that if  
\eq{
\label{lambdakest}
\Lambda_K(\phi)\geq c\phi,\;\;\;\;\mbox{for some $c>0$}
}
in the vicinity of zero, then $S_N$ is bounded from below by a 
logarithmic growth. We will prove that \eqref{lambdakest} holds 
for any measurable set $K\subset[0,1)$ (apart from the trivial cases, 
where $\Lambda_K(\phi)=0$). 

It is known from Lebesgue density theorem that for any measurable set 
$K$, $|K|=|K^d|$ holds, where $K^d$ denotes the set of the density points of 
$K$:
\[
K^d=\left\{x\in K\Biggm| \lim_{\delta\to 0}\frac{(x-\delta,x+\delta)
\cap K}{2\delta}=1\right\}.
\]
It can be inferred from this theorem that for any $K$ of positive measure, 
there is such a point $x\in K$ that
\eq{
\label{densp}
\begin{split}
\forall\epsilon>0:\;\exists\delta>0\;\;&
\mbox{so that for every interval $I$ 
that satisfies $x\in I$, and $|I|<\delta$,}\\ 
&|K\cap I|>(1-\epsilon)|I|.
\end{split}
}
Disregarding the trivial cases, the measure of $K^c$ (the complement of $K$), 
is also positive: 
$|K^c|>0$. It means that $K^c$ also has a point that satisfies \eqref{densp}. 
We denote this point by $y$. For a given $\epsilon$, we can chose a common 
$\delta$ to $x$ and $y$. Let $I$ be an interval shorter than this $\delta$: 
$|I|<\delta$, and $x\in I$. There is an integer $n$ such that $y\in(I+n|I|)$. 
The set $(I+n|I|)$ can be assured to be disjoint from $I$ by choosing a 
sufficiently small $\delta$. The following inequalities hold for $I$:
\eq{\label{iineq}
|K\cap I|>(1-\epsilon)|I|,\;\;\;\;\;|K^c\cap(I+n|I|)|>(1-\epsilon)|I|,} 
The following estimate, though seemingly weak, is the core of the proof:
\begin{eqnarray*}
\Lambda_K(|I|)&\geq&\left|\;\left(\bigcup_{k=0}^{n-1}(I+k|I|)\cap(K+|I|)\right)
\setminus K\;\right|\\
&=&\sum_{k=0}^{n-1}\big|\big((I+k|I|)\cap(K+|I|)\big)
\setminus\big(\left(I+(k+1)|I|\right)\cap K\big)\big|\\
&\geq&\sum_{k=0}^{n-1}\big|\big((I+k|I|)\cap (K+|I|)\big)\big|-\big|
\big(I+(k+1)|I|\big)\cap K\big|\\
&=&\big| I\cap K\big|-\big|(I+n|I|)\cap K\big| 
\end{eqnarray*}
Having a look at \eqref{iineq}, we obtain that for arbitrary 
positive $\epsilon$,
\[
\Lambda_K(|I|)\geq(1-2\epsilon)|I|,
\]
if $|I|$ is sufficiently small. This inequality entails \eqref{lambdakest}.  
\end{proof}

\subsection*{Conclusion and Outlook}

\hspace{\parindent}We have shown that for any sublinear 
growth $f_{N}$, there exist 
shift-invariant pure states that have faster entropy growths than  
$f_{N}$. However, the question if the entropy asymptotics of
any translation-invariant pure state is sublinear,
that is whether they have a vanishing entropy density, is still unsolved.
It is difficult to address this problem generally. One can instead take
into consideration only special classes of translation-invariant states.
For instance, finitely correlated states turn out to lead to bounded
entropy growth.\cite{Fin} A further step could be to explore the entropy
growths and the entropy densities of pure algebraic states, which are the 
generalizations of the finitely correlated states.   

Another question that can be raised is whether there exists for each 
sublinear growth $f_N$ a state with local von Neumann entropies $S_N$ such that
$\lim_{N\to \infty} S_N/f_N=c$, 
where $c > 0$. As we can learn from the last
proposition in Sec. III, in the case of pure translation-invariant 
quasifree state the answer is negative.  

In the case of local Hamiltonians only ground states with bounded or
logarithmic entropy growth have been found. From a mathematical point of
view, our construction is not sophisticated. Any given sublinear asymptotics
is exceedable by the entropy growth of a state characterized by a set of rather
simple structure: a set built from countably many intervals. Nevertheless, it 
is still an open question whether these asymptotics can be physically 
realized, or entropy asymptotics stronger than logarithmic (or some other 
sublinear) function can never occur
for ground states in the presence of only local interactions.

\subsection*{Acknowledgments}

\hspace{\parindent}We would like to thank M. Mosonyi 
and P. Vecserny\'es for helpful 
discussions and thorough reading of the manuscript.
One of the authors (Z. Z.) was partially supported by OTKA Grant No. 
T043159.

\end{document}